\documentclass{article}
\usepackage[T1]{fontenc} 
\usepackage[utf8]{inputenc} 
\usepackage{ismir,amsmath,cite,url}
\usepackage{graphicx}
\usepackage{color}
\usepackage{multirow}
\usepackage{multicol}
\usepackage{subfigure}
\usepackage{amsfonts}
\usepackage{verbatim}
\usepackage{float}
\usepackage{lineno}

\title{A unified model for zero-shot music source separation, transcription and synthesis}

\multauthor
{Liwei Lin$^1$ \hspace{1cm} Qiuqiang Kong$^2$
\hspace{1cm} Junyan Jiang$^1$
 \hspace{1cm} Gus Xia$^1$} { 
  $^1$ Music X Lab, New York University Shanghai\\
$^2$ ByteDance, Shanghai, China\\
{\tt\small linliwei3916@gmail.com,  kongqiuqiang@bytedance.com,jj2732@nyu.edu,gxia@nyu.edu}
}

\def\authorname{Liwei Lin, Qiuqiang Kong, Junyan Jiang and Gus Xia}

\begin{document}

\maketitle
\begin{abstract}

We propose a unified model for three inter-related tasks: 1) to \textit{separate} individual sound sources from a mixed music audio, 2) to \textit{transcribe} each sound source to MIDI notes, and 3) to\textit{ synthesize} new pieces based on the timbre of separated sources. The model is inspired by the fact that when humans listen to music, our minds can not only separate the sounds of different instruments, but also at the same time perceive high-level representations such as score and timbre. To mirror such capability computationally, we designed a pitch-timbre disentanglement module based on a popular encoder-decoder neural architecture for source separation. The key inductive biases are vector-quantization for pitch representation and pitch-transformation invariant for timbre representation. In addition, we adopted a query-by-example method to achieve \textit{zero-shot} learning, i.e., the model is capable of doing source separation, transcription, and synthesis for \textit{unseen} instruments. The current design focuses on audio mixtures of two monophonic instruments. Experimental results show that our model outperforms existing multi-task baselines, and the transcribed score serves as a powerful auxiliary for separation tasks.

\end{abstract}
\section{Introduction}\label{sec:introduction}

Music source separation (MSS) is a core problem in music information retrieval (MIR), which aims to separate individual sound sources, either instrumental or vocal, from a mixed music audio. A good separation benefits various of downstream tasks of music understanding and generation\cite{yoshioka2018multi,8462081} since many music-processing algorithms call for ``clean'' sound sources.


With the development of deep neural networks, we see significant performance improvements in MSS. The current mainstream methodology is to train on pre-defined music sources and then infer a mask on the spectrogram (or other data representations) of the mixed audio. More recently, we see several new efforts in MSS research, including query-based method \cite{QueryLee2019audio,ClassSeetharaman2019class,MetaSamuel2020meta,HieraISMIR2020Manilowhierarchical} for unseen (not pre-defined) sources, semantic-based separation that incorporates auxiliary information such as score or video\cite{woodruff2006remixing,miron2017monaural,ewert2017structured,Gover2019ScoreInformedSS,meseguerbrocal2020content,VisualZhao2018sound,VisualGan2020music}, and multi-task settings\cite{SimultaneousManilow2020simultaneous}. 


This study conceptually combines the aforementioned new ideas but follows a very different methodology --- instead of directly applying masks, \textit{we regard MSS an audio pitch-timbre disentanglement and reconstruction problem}. Such strategy is inspired  by the fact that when humans listen to music, our minds not only separate the sounds into different sources but also perceive high-level pitch and timbre representations that generalize well during both music understanding and creation. For example, humans can easily identify the same timbre in other pieces or identify the same piece played by other instruments. People can even mimic the learned timbre using human voice and sing (i.e., to synthesize via voice) the learned pitch sequence.

To mirror such capability computationally, we propose a zero-shot multi-task model jointly performing MSS, automatic music transcription (AMT), and synthesis. The model comprises four components: 1) a query-by-example (QBE) network, 2) a pitch-timbre disentanglement module, 3) a transcriptor, and 4) an audio encoder-decoder network. First, the QBE network summarizes the clean query example audio (which contains only one instrument) into a low-dimensional query vector, conditioned on which the audio encoder extracts the latent representation of an individual sound source. Second, the model disentangles the latent representation into pitch and timbre vectors while transcribing the score 
using the transcriptor.
Finally, the audio decoder takes in both the disentangled pitch and timbre representations, generating a separated sound source. When the model further equips the timbre representation with a pitch-transformation invariance loss, the decoder becomes a synthesizer, capable of generating new sounds based on an existing timbre vector and new scores.

The current model focuses on audio mixtures of two monophonic instruments and performs in a frame-by-frame fashion. Also, it only transcribes pitch and duration information. We leave polyphonic and vocal scenarios as well as a more complete transcription for future work. In sum, our contributions are:

\begin{itemize}
\item \textbf{Zero-shot multi-task modeling}: To the best of our knowledge, it is the first model that jointly performs separation, transcription, and synthesis. It works for both previously seen and unseen sources using a query-based method.
\item \textbf{Well-suited inductive bias}: The neural structure is analogous to the ``hardware'' of the model, which alone is inadequate to achieve good disentanglement. We designed  two extra inductive biases: vector-quantization for pitch representation and pitch-transformation invariant for timbre representation, which serves as a critical part of the ``software'' of the model. 
\item  \textbf{None-mask-based MSS}: Our methodology regards MSS an audio pitch-timbre disentanglement and re-creation problem, unifying music understanding and generation in a representation learning framework.
\end{itemize}

\section{Related work}\label{related work}

Most effective music source separation (MSS) methods are based on well-designed neural networks, such as U-Net\cite{DUnetJansson2017orcid} and MMDenseLSTM\cite{takahashi2018mmdenselstm}. Here, we review three new trends of MSS related of our work:  1) multi-task learning, 2) zero-shot for unseen sources, and 3) taking advantage of auxiliary semantic information.

\subsection{Multi-task Separation and Transcription}
Several recent studies\cite{SimultaneousManilow2020simultaneous,MultitaskHung2020multitask,DPMtanakamulti} conduct multi-task separation and transcription by learning a joint representation for both tasks. These works demonstrated that a multi-task setting benefits one or both of the two tasks due to the better generalized capability of the learned joint representation. Our model is a multi-task and can further disentangle pitch and timbre representation for sound synthesis.

\begin{figure*}[t]
  \centering
  \vskip -0.1in
  \subfigure[Baseline models]{\label{fig1a}\includegraphics[width=\columnwidth]{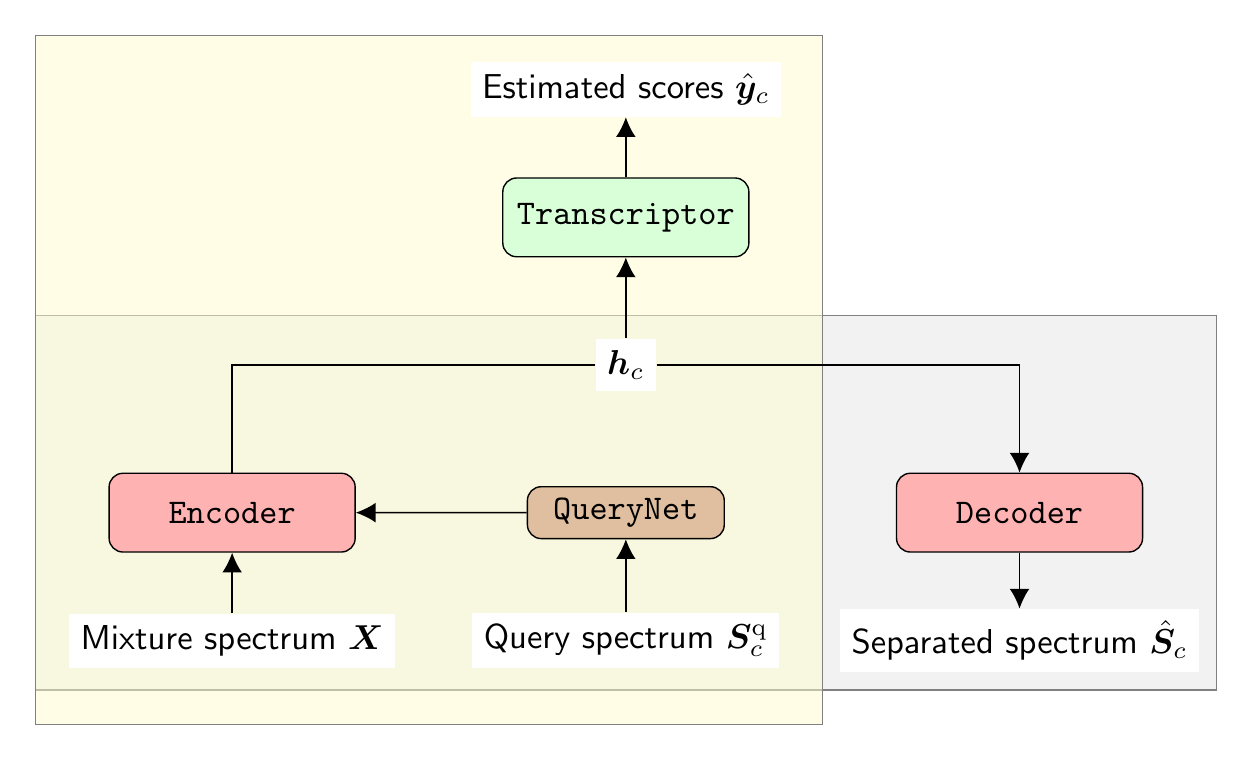}}
  \subfigure[The proposed model]{\label{fig1b}\includegraphics[width=\columnwidth]{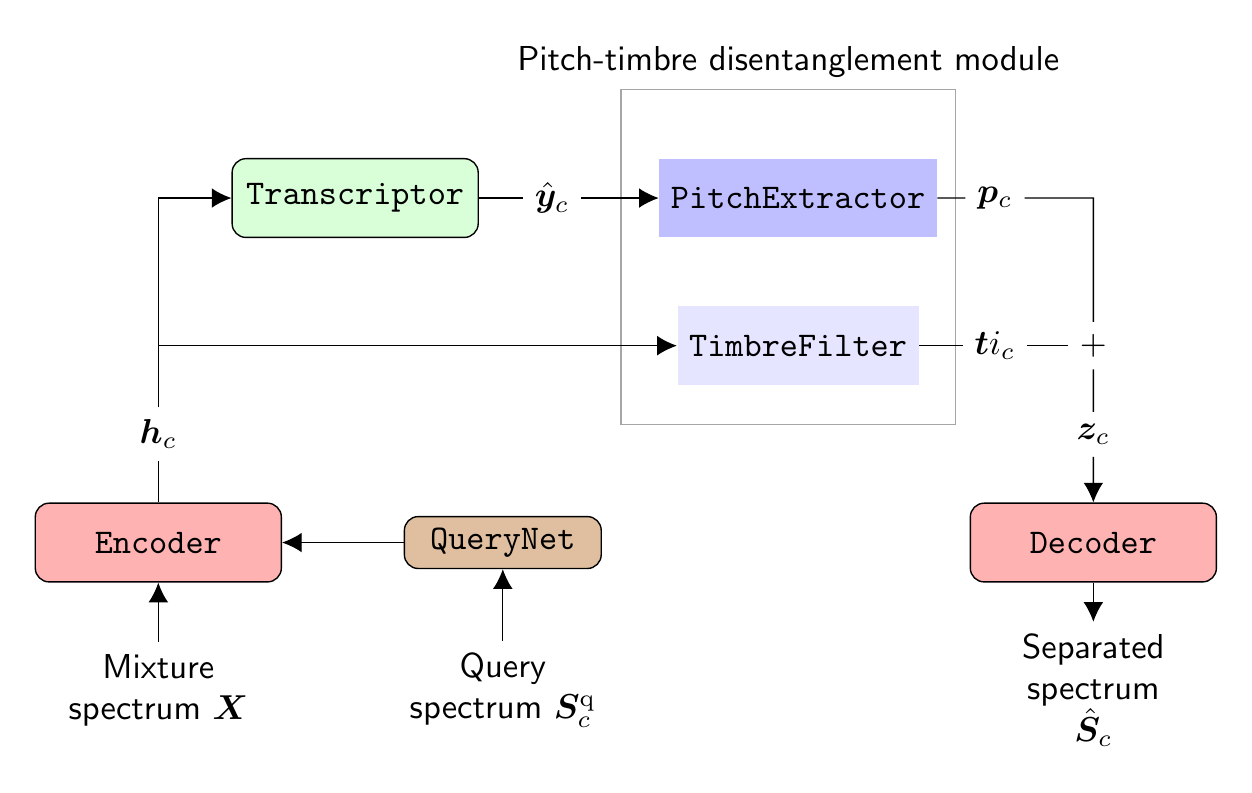}}
  \caption{The baseline models and the proposed model. In the left figure, the large orange and gray box indicate a QBE transcription-only and QBE separation-only model respectively. The whole figure indicates a QBE multi-task model.}
  \label{fig:workflow}
\end{figure*}
\subsection{Query-based Separation}\label{Query-based Separation}

Few-shot and zero-shot learning are becoming popular in MIR. For the MSS task, It is meaningful to separate unseen rather than pre-defined sources since it is unrealistic to collect training data that covers all the sources with considerable amounts. Query-by-example (QBE) network is one of solutions for zero-shot learning and recent researches \cite{QueryLee2019audio,ClassSeetharaman2019class,DChershey2016deep,DCwang2018alternative,ATluo2018speaker,ATkumar2018music,HieraISMIR2020Manilowhierarchical} show its nice performance. In this study, we adopt a QBE as in \cite{QueryLee2019audio}.

\subsection{Semantic-based Separation}\label{Content-conditioned Separation}
Many researches demonstrate that semantic information is a useful auxiliary for MSS. For example, Gover et al.\cite{Gover2019ScoreInformedSS} designs a score-informed wave-U-Net to separate choral music; Jeon et al.\cite{LyricsJeon2020exploring} performs the lyrics-informed separation; Meseguer-Brocal et al.\cite{meseguerbrocal2020content} develops a phoneme-informed C-U-Net\cite{meseguer2019conditioned}; Zhao et al.\cite{VisualZhao2018sound} takes advantage of visual information to separate homogeneous instruments. But these methods cannot separate sources without additional semantic groundtruths during inference. Our study can also be regarded as score-informed MSS, but our model does not call for ground truth score during the inference time.


\section{Methodology}\label{methodology}
In this section, we describe our proposed 1) multi-task and QBE model for source separation; 2) pitch-timbre disentanglement module; 3) pitch-translation invariance loss.


\subsection{Multi-task Separation and Transcription}
\label{Multi-task separation and transcription}



Different from previous works that tackle the music separation and music transcription problems separately, we learn a joint representation for both of them. Previous works \cite{SimultaneousManilow2020simultaneous,MultitaskHung2020multitask} have shown that the representation learnt by a joint separation and transcription task can generalize better than the representation learnt by single-task models.

We denote the waveform of two single-source audio segments from different sources as $ {\bm s}_{c} \in \mathbb{R}^{L} $ and $ {\bm s}_{i} \in \mathbb{R}^{L} $, respectively.
We denote their mixture as:
\begin{equation}
   {\bm x}= {\bm s}_{c} +{\bm s}_{i}. 
   \label{eq0}
\end{equation}
\noindent Our aim is to separate $ {\bm s}_{c} $ from $ {\bm x} $.
We denote the spectrogram of $ {\bm x} $ and $ {\bm s}_{c} $ as $\bm{X}\in {\mathbb R}^{T\times F}$ and ${\bm{S}}_c \in {\mathbb R}^{T\times F}$, respectively.

We first formalize the general MSS model using an encoder-decoder neural architecture. For instance, UNet \cite{DUnetJansson2017orcid} is an encoder-decoder architecture which is widely used in MSS. By ignoring the skip connections of UNet, the output of the encoder (the bottleneck of U-Net) can be used as a joint representation for separation and transcription. Different from previous MSS methods that estimate a single-target mask on the mixture spectrogram, we design the separation model to directly output spectrograms. In this way, the model can not only separate a source from a mixture, but can also synthesize new audio recordings from joint representations.

 


For the source separation system, we denote the encoder and decoder as follows:

\begin{equation}
\bm h = {\rm Encoder}(\bm{X})
\label{eq1},
\end{equation}
\begin{equation}
\widehat{\bm S}_c = {\rm Decoder}_c(\bm h)
\label{eq2},
\end{equation}
where $\bm h$ is the learned joint representation and ${\rm Decoder}_c$ is the decoder for the target source $c$. The joint representation $\bm h$ is used as input to the a transcription model:
\begin{equation}
\widehat{\bm y}_c = {\rm Transcriptor}_c(\bm h),
\label{eq3}
\end{equation}
where $\widehat{\bm y}_c \in [0,1]^{T \times N}$ are probabilities of the predicted MIDI roll. Typically, we set $N=89$ including $88$ notes on a piano and a silence state.

When designing neural networks, to remain the transcription resolution, we do not apply temporal pooling operation in the encoder, decoder and transcriptor. So that the temporal resolution of $\bm h$ is consistent with that of $\bm{X}$. We describe the details of the encoder, decoder, and transcriptor in Section~\ref{section:model}.


\subsection{Query-by-example Separation and Transcription}
\label{Query-by-example separation and transcription}
As described in Equation~(\ref{eq2}) and (\ref{eq3}), we need to build $ J $ decoders to separate $ J $ target sources. With the number of target sources increases, the parameters number will also increase. More importantly, the model trained for pre-defined sources can not adapt to unseen sources. To tackle these problems, we design a QBE module in our model. The advantage of using QBE is that we can separate unseen target sources. That is, we achieve a zero-shot separation.

Similar to the QueryNet \cite{QueryLee2019audio}, we design a QueryNet module as shown in Figure~\ref{fig1a}. The QueryNet module extracts the embedding vector ${\bm q}_c \in {\mathbb R}^{M}$ of an input spectrogram ${\bm S}^{\rm q}_c \in {\mathbb R}^{T\times F}$, where $ M$ is the dimension of the embedding vector:

\begin{equation}
{\bm q}_{c} = {\rm QueryNet}({\bm S}^{\rm q}_c).
\label{eq4}
\end{equation}
Audio recordings from the same source will be learnt to have similar embedding vectors. We propose a contrastive loss to encourage embedding vectors from the same sources to be close, and embedding vectors from different sources to be far:
\begin{equation}
L_{\rm query} = \frac{1}{C} \{{\left \| {\bm q}_{c} -{\bm q}^{'}_{c} \right \|}_2 + \sum_{j \neq c}^{C} \max (m - {\left \| { \bm q}_{c} -{\bm q}_{j}\right \|}_2, 0)\},
\label{eq5}
\end{equation}
where $m>0$ is a margin and ${\bm q}^{'}_{c}$ and ${\bm q}_{c}$ are from the same source $c$, and ${\bm q}_{c}$ and ${\bm q}_{j}$ are from different sources. We set $M$ to $6$ and $m$ to $0.125$ in our experiments.

We input the embedding vector ${\bm q}_{c}$ as a condition to each layer of the encoder using Feature-wise Linear Modulation (FiLM)\cite{perez2018film} layers. Then the encoder outputs a representation ${\bm h}_c \in {\mathbb R}^{T\times D}$. The embedding vector ${\bm q}_{c}$ controls what source to separate or transcribe. There are only one encoder, one decoder, and one transcriptor to separate any sources:

\begin{equation}
{\bm h}_c = {\rm Encoder}(\bm{X}, {\bm q}_c),
\label{eq6}
\end{equation}
\begin{equation}
\widehat{\bm S}_c = {\rm Decoder}({\bm h}_c),
\label{eq7}
\end{equation}
\begin{equation}
\widehat{\bm{y}}_c = {\rm Transcriptor}({\bm h}_c).
\label{eq8}
\end{equation}

\subsection{Pitch-timbre Disentanglement Module}
\label{Disentangled pitch and timbre representations}
Previous MSS works do not disentangle pitch and timbre for separation. That is, those MSS methods implement separation systems without estimating pitches. In this section, we propose a pitch-timbre disentanglement module based on the query-based encoder-decoder architecture described in previous sections to learn interpretable representations for MSS. Such interpretable representations enable the model to achieve score-informed separation based on predicted scores. 

As shown in Figure~\ref{fig1b}, the proposed pitch-timbre disentanglement module consists of a PitchExtractor and a TimbreFilter module. The output of PitchExtractor $ {\bm p}_c $ only contains pitch information of $ {\bm s}_c $, and the output of TimbreFitler is expected to only contain timbre information of $ {\bm s}_c $. The PitchExtractor is modeled by an embedding layer $\bm V=\{{\bm e}_1, {\bm e}_2,...,{\bm e}_N\}$, where $ N $ is the number of vectors, which equals to the number of pitches $89$ in our experiment. To explain, ${\bm e}_n \in \bm V ({\bm e}_n\in \mathbb{R}^{K})$ denotes the quantized pitch vector for the $n$-th MIDI note. Then, we calculate the disentangled pitch representation for $ \hat{\bm y} $ as $\bm p_{c} = [{\bm p}^{(1)}_c;{\bm p}^{(2)}_c;...;{\bm p}^{(T)}_c]$, where ${\bm p}^{(t)}_c \in \mathbb{R}^{K}$:
\begin{equation}
\label{eq:y_dot_V}
{\bm p}^{(t)}_c = \sum_{n}^{N} \hat{y}_c^{(t, n)}\cdot{\bm e}_n,
\end{equation}
where $\hat{y}_c^{(t, n)}$ is the output of the transcriptor containing predicted presence probability of the $n$-th MIDI note or the silence state at time $t$, and $K$ is the dimension of the disentangled pitch representation. During synthesis, we can replace $\hat{\bm y}_c$ with one-hot encodings of new scores as input to Equation~(\ref{eq:y_dot_V}) to obtain pitch representation $ {\bm p}_c $ for synthesizing audio recordings.

TimbreFilter is used to filter timbre information ${\bm {ti}}_c \in {\mathbb R}^{T\times D}$ from ${\bm h}_c$:
\begin{equation}
{\bm {ti}}_c = {\rm TimbreFilter}({\bm h}_c).
\label{eq9}
\end{equation}
Here, TimbreFilter is modeled by a convolutional neural network. 
\noindent Then, we can synthesize $\hat{\bm s}_c$ using disentangled pitch ${\bm p}_c$ and timbre ${\bm {ti}}_c$. Inspired by the FiLM \cite{perez2018film}, we first split ${\bm p}_c$ into $ {\bm p}^{\gamma}_c $ and $ {\bm p}^{\beta}_c $, where $ {\bm p}_c = [{\bm p}^{\gamma}_c\, {\bm p}^{\beta}_c] $. Then, we entangle ${\bm p}_c$ and ${\bm {ti}}_c$ together to produce $\widehat{\bm S}_c$:
\begin{equation}
{\bm z}_c = {\bm p}^{\gamma}_c\odot {\bm {ti}}_c + {\bm p}^{\beta}_c,
\label{eq13}
\end{equation}
\begin{equation}
\widehat{\bm S}_c = {\rm Decoder}({\bm z}_c),
\label{eq14}
\end{equation}
and the separation loss is:
\begin{equation}
L_{\rm separation} ={ \left \| {\bm S}_c  - \widehat{\bm S}_c \right \|}_1.
\label{eq15}
\end{equation}
Different from previous MSS works, we apply a separation loss and a transcription loss to train the proposed model. The transcription loss is:

\begin{equation}
L_{\rm transcription} = \rm Cross\_entropy({\bm y}_c, \widehat{\bm y}_c).
\end{equation}
Here, ${\bm y}_c \in [0, 1]^{T\times N}$ is the groundtruth of scores.
The aggregated loss function is:
\begin{equation}
L = L_{\rm query} + L_{\rm transcription} + L_{\rm separation}.
\label{eq16}
\end{equation}

\indent The aggregated loss $L$ drives the proposed model to be a \emph{multi-task score-informed} model rather than a synthesizer due to the lack of inductive biases for further timbre disentanglement.

\subsection{Pitch-translation Invariance Loss}
\label{Pitch transformation invariance loss}

We propose a pitch-translation invariance loss to further improve the timbre disentanglement performance. Based on the pitch-translation invariance, we assume that when the audio pitches with the corresponding MIDI is shifted within a certain interval, the timbre is unchanged.

We shift the pitch of ${\bm s}_{c}$ to generate an augmented audio ${\bm s}_{c}^{'}$. The augmented audio ${\bm s}_{c}^{'}$ has the same timbre as ${\bm s}_{c}$. According to Equation~(\ref{eq0}), we have a new mixture audio ${\bm x}^{'}$:
\begin{equation}
    {\bm x}^{'}= {\bm s}_{c}^{'} +{\bm s}_{i}.
\label{eq18}
\end{equation}

\noindent We denote the ${\bm X}^{'}$ and ${\bm S}^{'}_c$ as the spectrograms of ${\bm x}^{'}$ and ${\bm s}_{c}^{'}$ respectively. We extract the disentangled timbre vector of ${\bm s}_{c}$, and denote it as ${\bm {ti}}^{'}_c$. Because ${\bm s}^{'}_c$ is pitch shifted ${\bm s}_c$, so that the timbre ${\bm {ti}}^{'}_c$ should be consistent with that of ${\bm {ti}}_c$. Therefore,the reconstructed spectrogram by the timbre ${\bm {ti}}^{'}_c$ and the pitch ${\bm p}_c$ should be consistent with ${\bm S}_c$:

\begin{equation}
\widehat{\bm z}^{'}_c = {\bm p}^{\gamma}_c\odot {\bm {ti}}^{'}_c + {\bm p}^{\beta}_c,
\label{eq19}
\end{equation}
\begin{equation}
\widehat{\bm S}_c^{''} = {\rm Decoder}({\bm z}^{'}_c),
\label{eq20}
\end{equation}
\begin{equation}
    L_{\rm PTI} ={ \left \| {\bm S}_c  - \widehat{\bm S}_c^{''} \right \|}_1.
\label{eq21}
\end{equation}
\noindent where $ {\bm S}_c^{''} $ is the reconstructed spectrogram. We denote $L_{\rm PTI}$ as a pitch-translation invariance loss. With $L_{\rm PTI}$, our proposed model is capable of learning the disentanglement of pitch and timbre. A byproduct of the disentanglement system is that, the decoder of our system becomes a \emph{synthesizer}, which can be used to synthesize audio recordings using timbre and pitches as input. When we change $\widehat{\bm y}_c$ to arbitrary scores, our model can synthesis a new piece of music with the timbre of ${\bm S}_c$.

In total, the objective function we exploit to train the proposed model with further disentanglement includes a QueryNet loss, a transcription loss, and a pitch-translation invariance loss:
\begin{equation}
L^{'} = L_{\rm query} + L_{\rm transcription} + L_{\rm PTI}.
\label{eq22}
\end{equation}

\section{Experiments}
\subsection{Dataset and Pre-processing}
\begin{figure}
 \centerline{
 \includegraphics[width=\columnwidth]{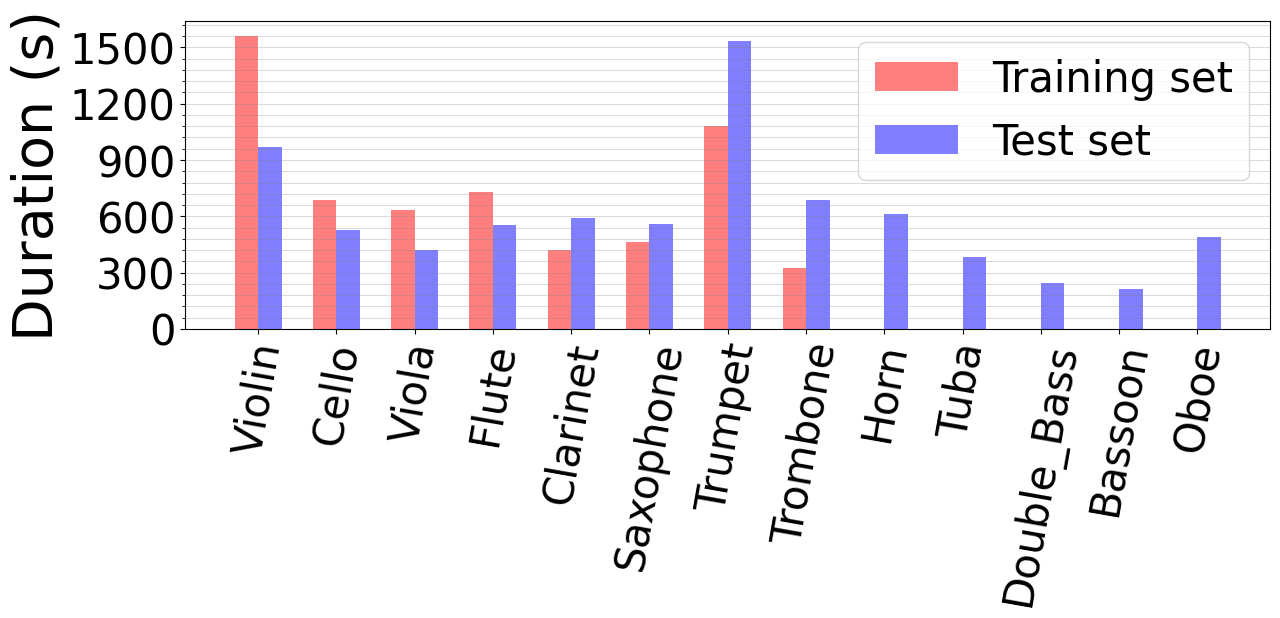}}
  \vskip -0.1in
 \caption{Duration of each instrument in the dataset.}
 \label{fig:data_distribution}
\end{figure}

We utilize the University of Rochester Multimodal Music Performance (URMP) dataset\cite{li2018creating} as the experimental dataset. The URMP dataset is a multi-instrument audio-visual dataset covering 44 classical chamber music pieces remixed from 115 single-source tracks of 13 different \emph{monophonic} instruments. The dataset provides note annotations for each single track. As shown in Figure~\ref{fig:data_distribution}, we divide these instruments into two groups (8 seen and 5 unseen instruments) and tracks into two sub-sets (55 tracks of 8 seen instruments for training and 32 songs by remixing 60 tracks of 13 instruments for test). Note that we calculate the duration of repeated tracks of different songs in the test set and do not exclude silence segments of all the tracks.

We resample all the tracks with a sample rate of 16KHz and extract them into Short-time Fourier transform (STFT) spectrograms with a window size of 1024 and 10ms overlap $(n_{\rm FFT}=2048)$. During training, we randomly remix \emph{2 arbitrary clips of different instruments} to generate a mixture. All the training data are augmented using pitch shifting ($\pm 4$ semitones) mentioned in Section~\ref{Pitch transformation invariance loss}. 

\begin{figure*}[t]
  \centering
   \vskip -0.1in
  \centerline{
 \includegraphics[width=2.1\columnwidth]{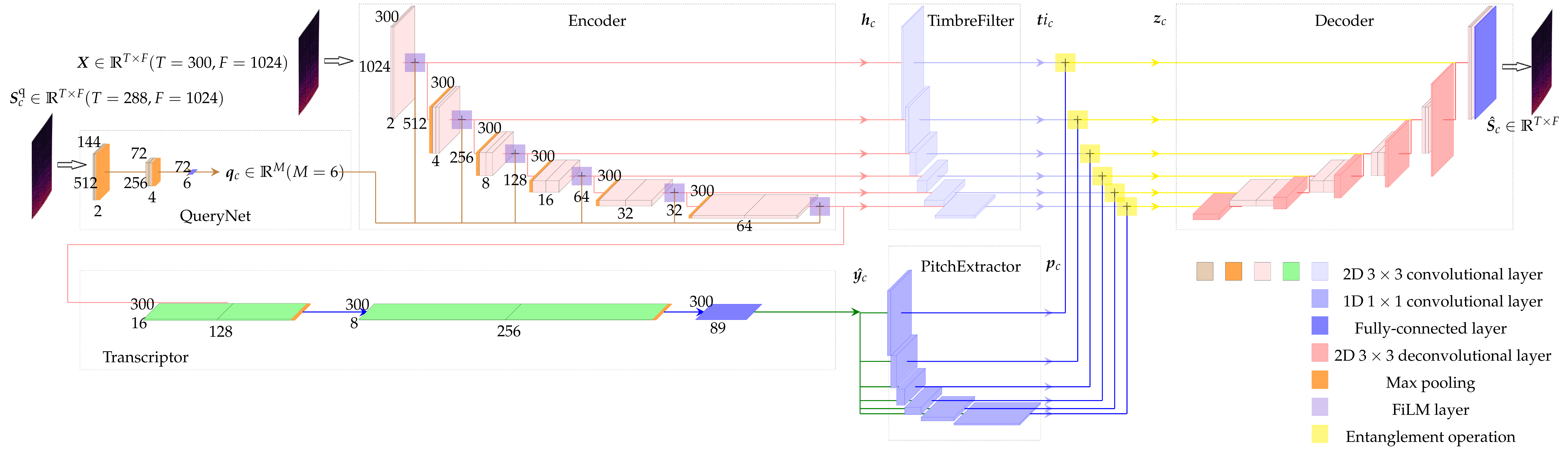}}
 \vskip -0.1in
  \caption{The model architecture with detailed hyper-parameter configuration.}
  \label{fig:model}
\end{figure*}

\begin{table*}[t]
\renewcommand{\arraystretch}{1.3}
 \begin{center}
 \setlength{
\tabcolsep}{0.7mm}{
 \begin{tabular}{c|cccc|cccc}
\hline
&\multicolumn{4}{c|}{\textbf{Separation (SDR)}}&\multicolumn{4}{c}{\textbf{Transcription (Precision)}}\\
  \hline
    \textbf{Model}
    &\textbf{MSS-only}
     &\textbf{Multi-task}
    &\textbf{MSI(ours)}
    &\textbf{MSI-DIS(ours)}
    &\textbf{AMT-only}
    &\textbf{Multi-task}
    &\textbf{MSI(ours)}
    &\textbf{MSI-DIS(ours)}
     \\
  \hline
  
\textbf{Seen}
&$4.69\pm0.31$
&$3.32\pm0.18$
&$\bm{6.33\pm0.17}$
&$5.04\pm0.16$
&$\bm{0.72\pm0.01}$
&$\bm{0.72\pm0.04}$
&$0.71\pm0.01$
&$\bm{0.72\pm0.01}$

\\
\textbf{Unseen}
&$\bm{6.20\pm0.26}$
&$4.63\pm0.34$
&$5.53\pm0.15$
&$3.99\pm0.22$
&$\bm{0.61\pm0.01}$
&$0.58\pm0.02$
&$\bm{0.61\pm0.01}$
&$0.59\pm0.01$

\\
  \textbf{Overall} 
&$5.07\pm0.22$
&$3.65\pm0.22$
&$\bm{6.13\pm0.15}$
&$4.77\pm0.14$
&$\bm{0.69\pm0.01}$
&$\bm{0.69\pm0.03}$
&$\bm{0.69\pm0.01}$
&$0.68\pm0.00$

\\
\hline
 \end{tabular}}
\end{center}
\vskip -0.1in
 \caption{The separation and transcription performance of all models..}
 \label{tab:exp_results}
\end{table*}
\begin{figure*}[!htp]
 \centerline{
 \includegraphics[width=2.1\columnwidth]{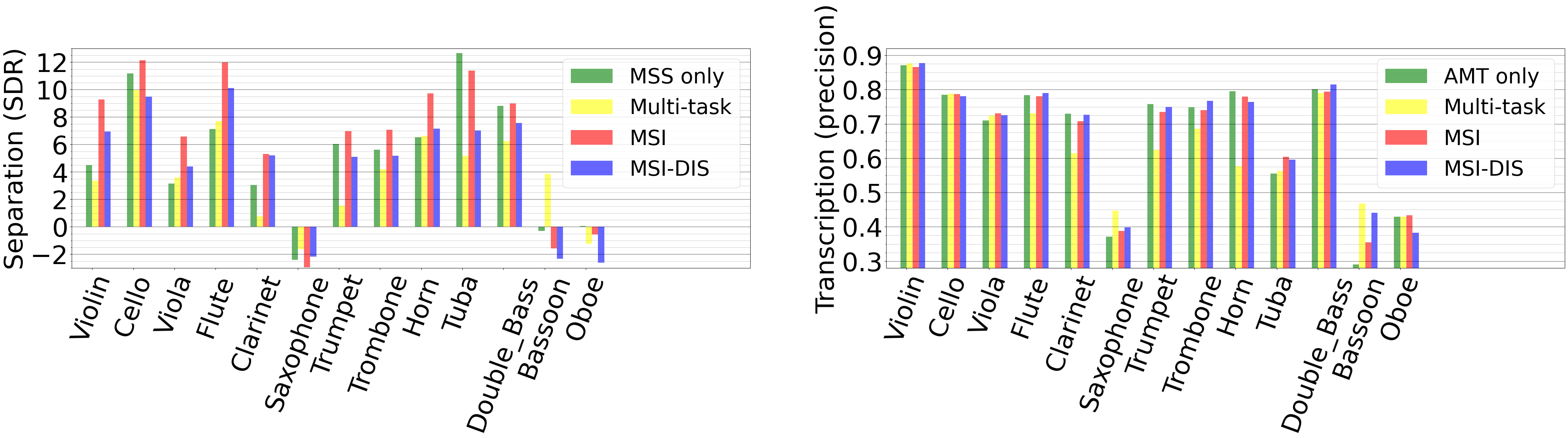}}
 \vskip -0.1in
 \caption{Instrument-wise performance of models. The last 5 instruments are unseen in the training set.}
 \label{fig:isntr_wise}
\end{figure*}
\subsection{Model Architecture}
\label{section:model}
We design our models based on U-Net, the current prominent model in MSS. Figure~\ref{fig1b} and~\ref{fig:model} elaborates details of the proposed multi-task score-informed model (MSI) described in Section~\ref{Disentangled pitch and timbre representations} and model with further disentanglement (MSI-DIS) illustrated in Section~\ref{Pitch transformation invariance loss}. 

\subsubsection{The Architecture of the MSI and MSI-DIS Model}
The combination of the encoder and decoder is a general U-Net without temporal pooling. The QueryNet comprises 2 CNN blocks, each of which consists of 2 convolution layers and a $2 \times 2$ max pooling module. A fully-connected layer and a tanh activation layer are applied to the last feature maps. We then average output vectors over the temporal axis to get a 6-dimensional query embedding vector ${\bm q}_c$. The architecture of the transcriptor is similar to the QueryNet but without temporal pooling. Each blue block in TimbreFilter depicted in Figure~\ref{fig:model} is a 2-dimension convolutional layer, the shape of the tensor output by which is as same as that of the input tensor. Each deep blue block in PitchExtractor is a 1-dimension $1 \times 1$ convolutional layer. Typically, the bottleneck of U-Net is regarded as ${\bm h}_c$. However, when constructing disentangled timbre representations, we regard the set of concatenate residual tensors as ${\bm h}_c$ to avoid non-disentangled representations leaking into the decoder.

Note that the kernel size of each 2-dimension convolutional layer is $3 \times 3$ and each 2-dimension convolutional layer (excepting TimbreFilter) is followed by a ReLU activation layer and a Batch Normalization layer.

\subsubsection{Baseline Design}
As shown in Figure~\ref{fig1a}, besides the proposed models illustrated above, we also report the performance of 3 extra baseline models in our experimental results. The QBE transcription-only baseline model (AMT-only) is composed of the queryNet, encoder, and transcriptor; the QBE separation-only baseline model (MSS-only) is a general U-Net; the QBE multi-task baseline model is composed of a U-Net and a transcriptor. All the hyper-parameters of components in these models are consistent with those of corresponding components in our models.

\begin{figure*}[!htp]
  \centering
 \vskip -0.1in
  \subfigure[MSI (synthesis)]{\label{spec1}\includegraphics[width=0.565\columnwidth]{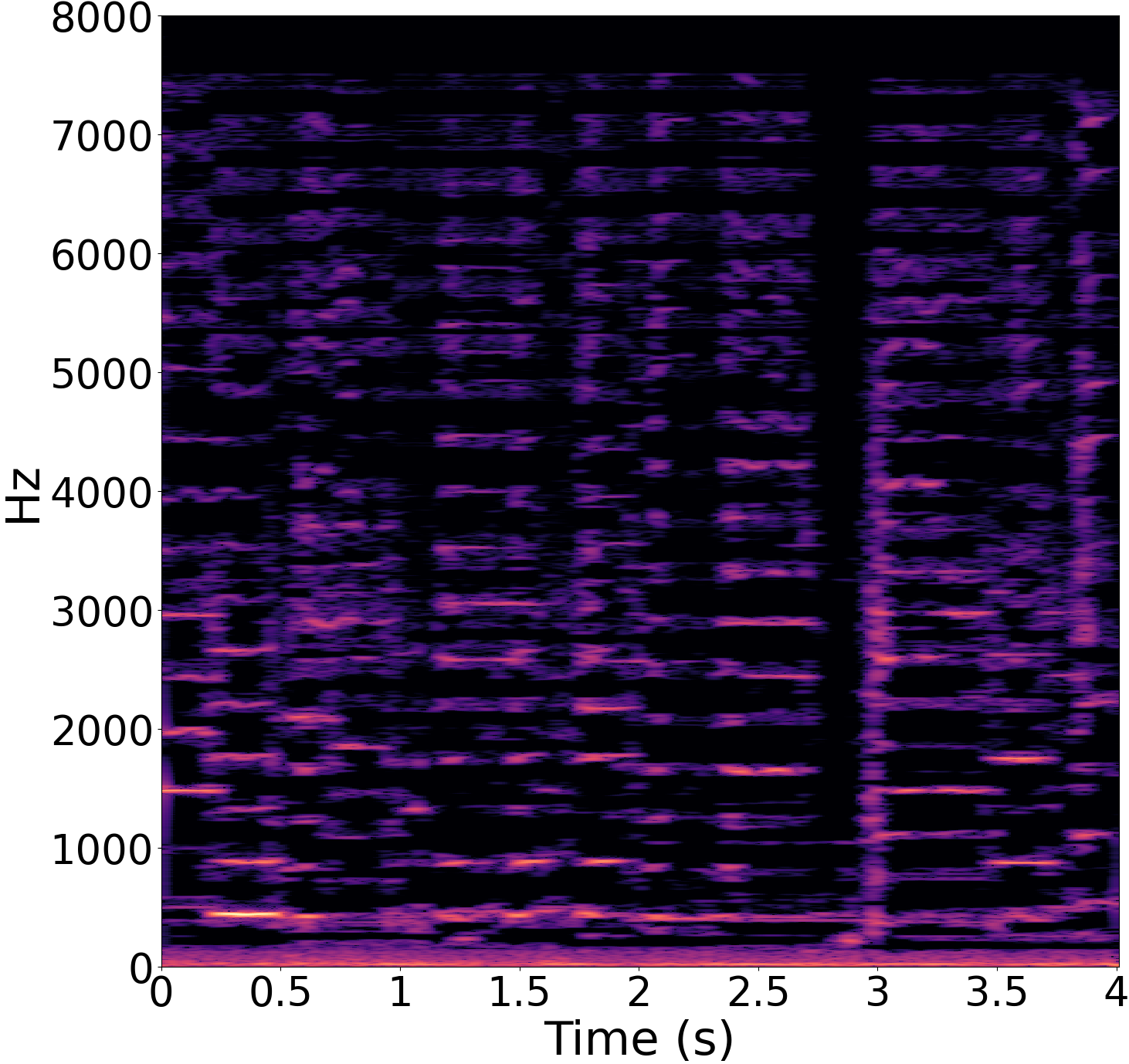}}
  \subfigure[MSI-DIS (synthesis)]{\label{spec2}\includegraphics[width=0.49\columnwidth]{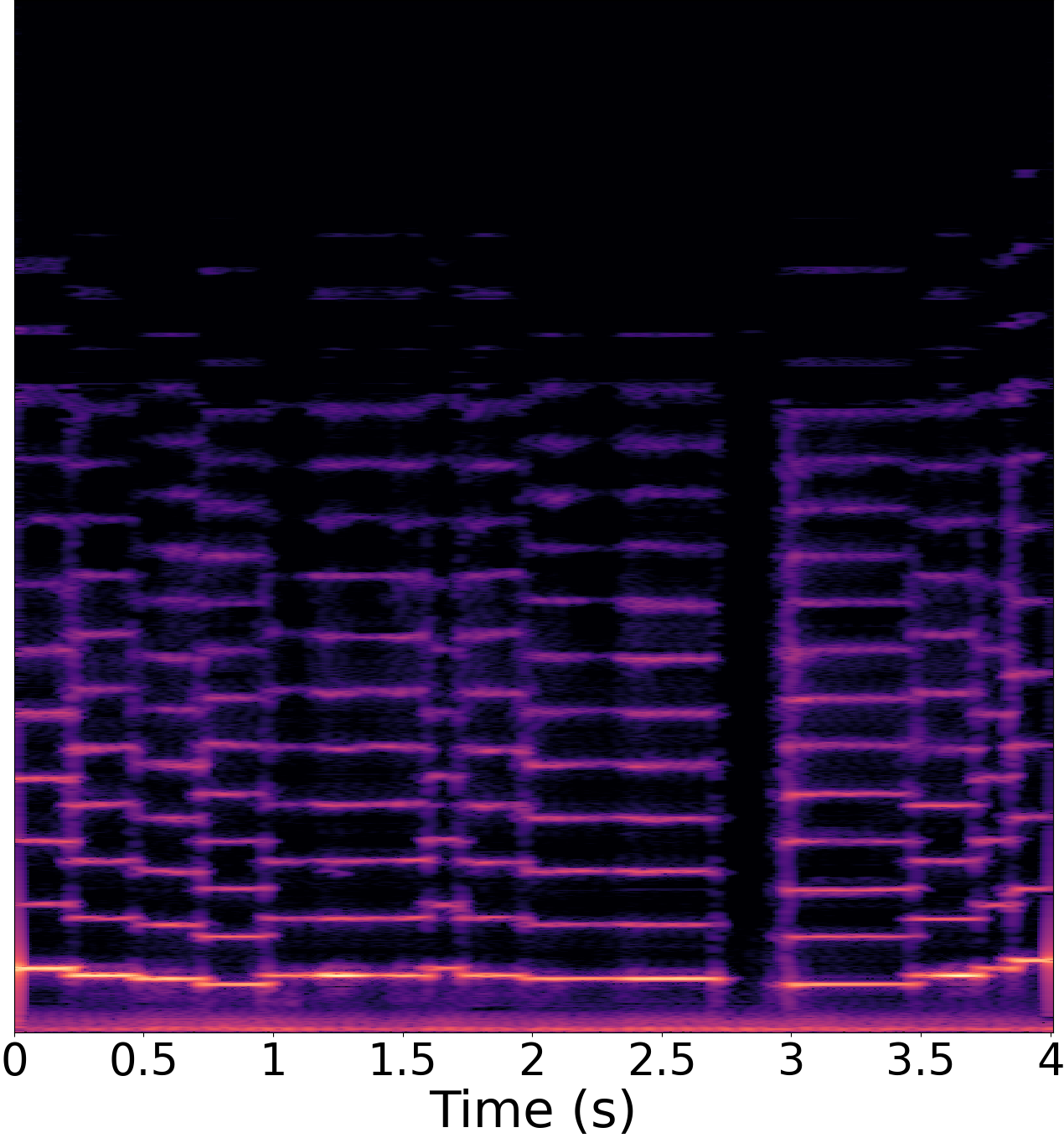}}
   \subfigure[MSI (separation)]{\label{spec3}\includegraphics[width=0.49\columnwidth]{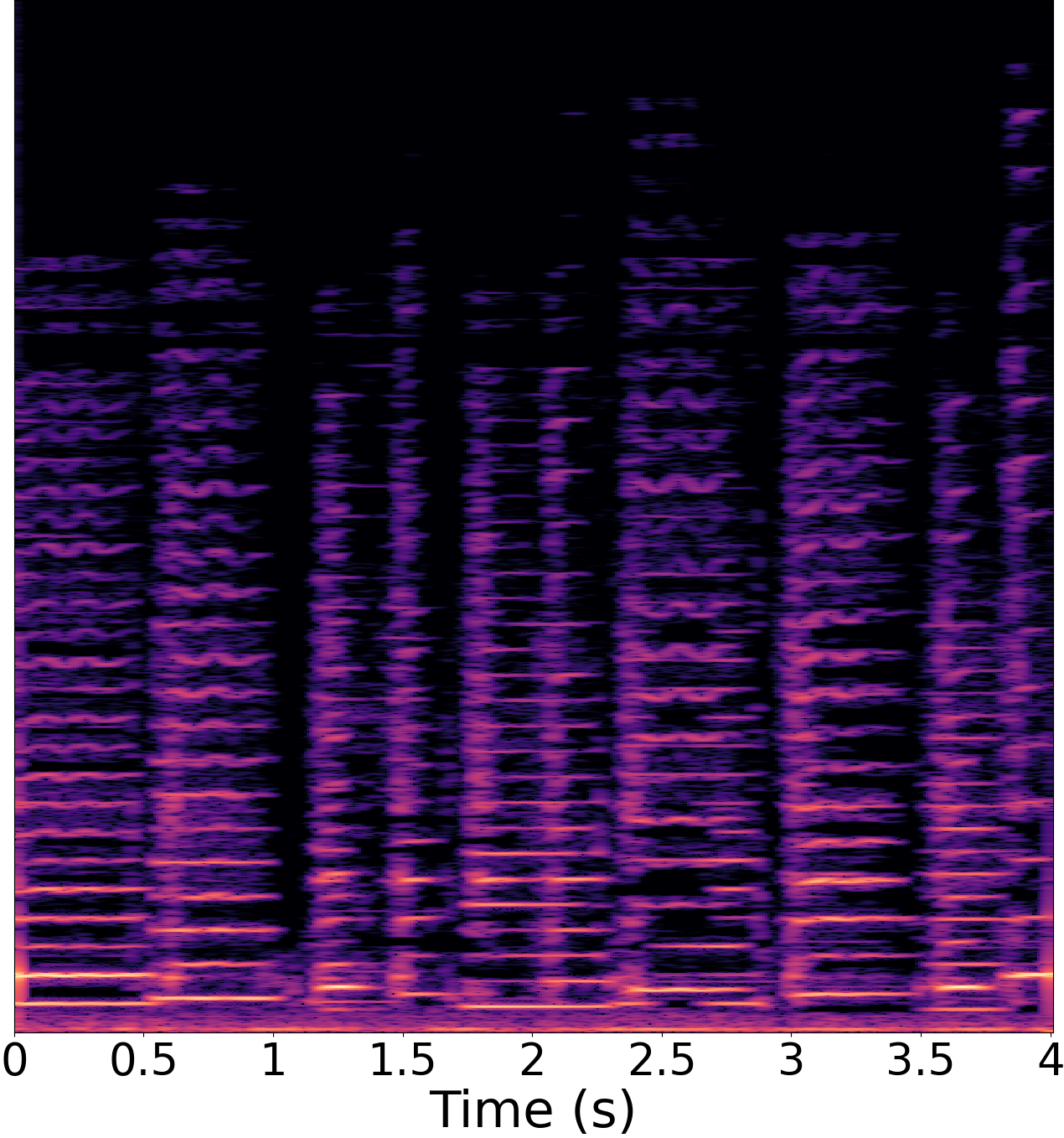}}
  \subfigure[MSI-DIS (separation)]{\label{spec4}\includegraphics[width=0.49\columnwidth]{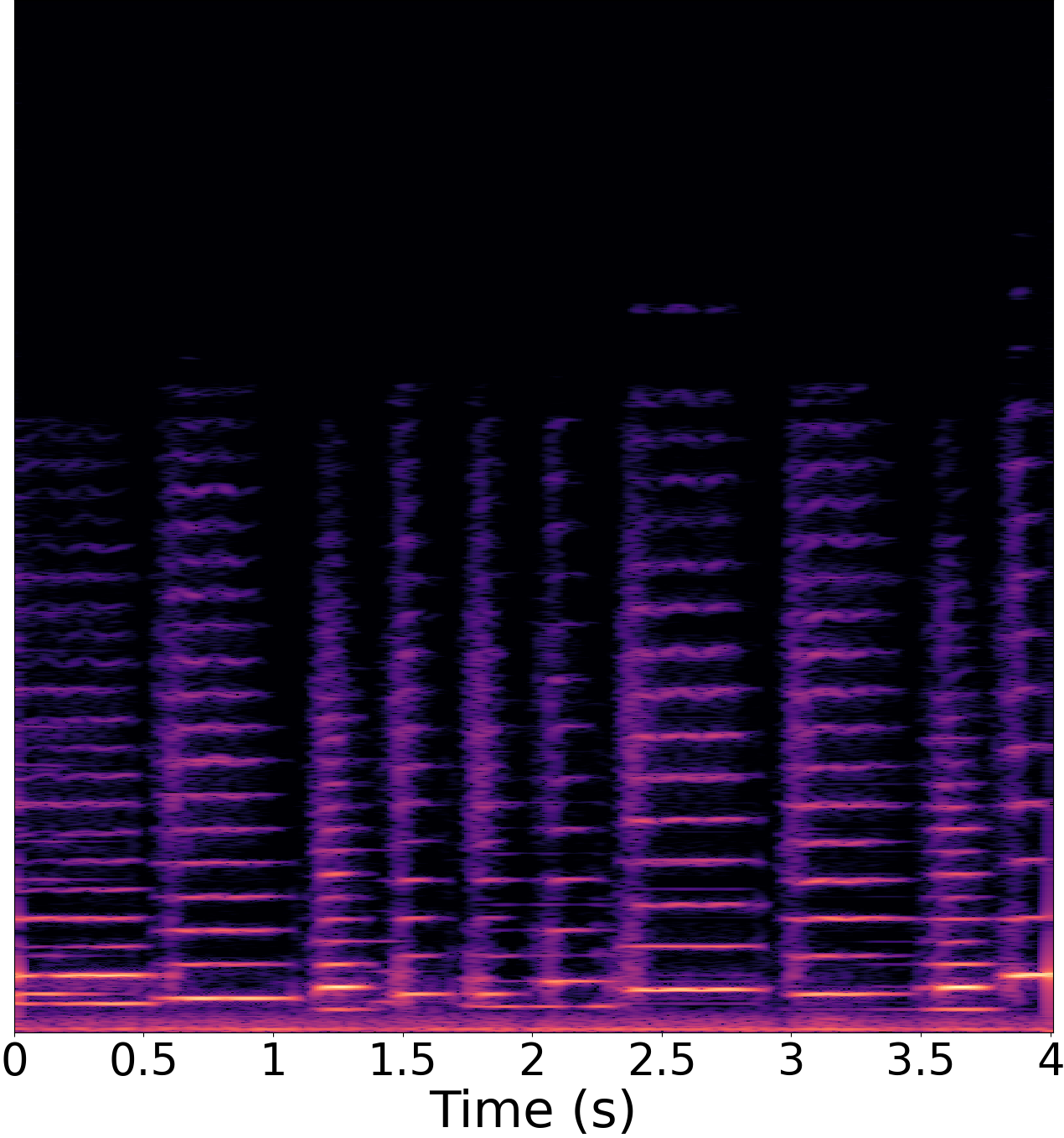}}
 \vskip -0.1in
  \caption{Spectrograms of audios synthesized and separated by the MSI and MSI-DIS model respectively. Models are expected to separate a viola source from the mixture of clarinet and viola. During synthesis, the two models are given the same new scores and are expected to synthesis new pieces with these scores and the separated viola timbre.}
  \label{fig:spec}
\end{figure*}
\subsection{Training and Evaluation}
All the models are trained with a mini-batch of 12 audio pairs for 200 epochs. All the models are evaluated with source-to-distortion (SDR) computed by mir\_eval pakage\cite{raffel2014mir_eval} for separation and precision computed by sklearn package\cite{scikit-learn} for transcription. During training, each audio pair comprises 2 single-track audio clips of different instruments to generate a mixture, 2 correspondings augmented samples for pitch-transformation invariance loss, and 3 single-track audio clips that exclude silence segments for contrastive loss. During inference, each test pair comprises a 4-second audio mixture and query sample. During synthesis, we employ Griffin Lim Algorithm (GLA)\cite{sharma2020fast} as the phase vocoder using torchaudio library. Since we do not divide a validation set to chose the best-performance model among all the training epochs, we report Micro-average results with a 95\% confidence interval (CI) of models at the last 10 epochs. All the experimental results are reproducible via our realeased source code\footnote{https://github.com/kikyo-16/a-unified-model-for-zero-shot-musical-source-separation-transcription-and-synthesis}.

\section{Results}

Experimental results shown in Table~\ref{tab:exp_results} demonstrate that the proposed MSI model outperforms baselines on separation without sacrificing performance on transcription.
The instrument-wise performance on unseen instruments depicted in Figure~\ref{fig:isntr_wise} demonstrates that the proposed models are capable of performing zero-shot transcription and separation. We also release synthesized audio demos online\footnote{https://kikyo-16.github.io/demo-page-of-a-unified-model-for-separation-transcriptiion-synthesis}. These demos demonstrate the success of the proposed inductive biases for disentanglement.

\subsection{Multi-task Baseline vs Single-task Baselines}
As shown in Table~\ref{tab:exp_results}, the multi-task baseline performs worse than the separation-only baseline, suggesting that the joint representation requires extra inductive biases to learn better generalization, i.e. deep clustering in Cerberus\cite{SimultaneousManilow2020simultaneous}. Our disentanglement strategy provides such inductive biases.

\subsection{MSI model vs Baselines}
With the auxiliary of the proposed pitch-timbre disentanglement module, compared with the multi-task baseline, the performance of MSI on separation becomes better. This indicates that the disentanglement module improve the generalization capability of the joint representation, leading to better separation results. Meanwhile, MSI outperforms the MSS-only baseline on separation by 1.06 points. This demonstrates that inaccurate scores transcribed by the model itself sever as a powerful auxiliary for separation.

\subsection{MSI Model vs MSI-DIS Model}
As depicted in Figure~\ref{spec1} and~\ref{spec2}, it is interesting that despite the same `'hardware'' (neural network design) of the two models, the MSI model fails to synthesis but the MSI-DIS model achieves. It exactly demonstrates that the designed `'soft-ware'' (the pitch translation loss) takes effect on the success of the disentanglement. As for separation performance shown in Table~\ref{tab:exp_results}, the MSI-DIS model falls behind the MSI model. The observation that better synthesis quality does not implies better separation performance suggests a trade-off between disentanglement and reconstruction. It indicates that extra (well-suited) inductive biases are required to further improve pitch and timbre disentanglement at the same time reduce the loss of information necessary for reconstruction.

Comparing the performance on seen with unseen instruments shown in Table~\ref{tab:exp_results}, we find that the separation quality of the MSI-DIS model is more sensitive to the accuracy of transcription results than that of the MSI model. This is because the MSI-DIS model synthesizes instead of separating sources, for which the separation performance of it relies more on the accuracy of transcription results and the capability of the decoder than the MSI model does. However, when comparing separated spectrograms shown in Figure~\ref{spec3} and~\ref{spec4}, we find that the MSI model sometimes separates multiple pitches at the same time while the MSI-DIS model yields monophonic results that sound more ``clean''. We release more synthesized and separated audio demos online.

\section{Conclusion and Future Works}
 We contributed a unified model for zero-shot music source separation, transcription, and synthesis via pitch and timbre disentanglement. The main novelty lies in the disentanglement-and-reconstruction methodology for source separation, which naturally empowers the model with transcription and synthesis capabilities. In addition, we designed well-suited inductive bias including pitch vector quantization and pitch-translation invariant timbre loss to achieve better disentanglement. Lastly, we successfully integrate the model with a query-based networks, so that all three tasks can be achieved in a zero-shot fashion for unseen sound sources. Experiments demonstrated the zero-shot capability of the model and the powerful auxiliary of disentangled pitch information to separation. Results of synthesized audio pieces further exhibit that the disentangled factors are well generalized. In the future, we plan to extent the proposed framework for multi-instrument and vocal scenarios as well as high-fidelity synthesis.
\bibliography{ISMIR2021_template}

%
%
%
%
%

\end{document}